\documentstyle[draft,epsf]{mn}

\def\gsim{\mathrel{\raise0.35ex\hbox{$\scriptstyle >$}\kern-0.6em 
\lower0.40ex\hbox{{$\scriptstyle \sim$}}}}
\def\lsim{\mathrel{\raise0.35ex\hbox{$\scriptstyle <$}\kern-0.6em 
\lower0.40ex\hbox{{$\scriptstyle \sim$}}}}

\title[Initial Mass Function of Low Surface Brightness Galaxies] 
{Is the Initial Mass Function of Low Surface Brightness Galaxies Dominated by Low Mass Stars?} 

\author[H.-c. Lee, B.K. Gibson, C. Flynn, D. Kawata, M.A. Beasley]
{Hyun-chul Lee$^1$$^,$$^2$, 
Brad K. Gibson$^1$, 
Chris Flynn$^1$$^,$$^3$, 
Daisuke Kawata$^1$ and 
\vspace{0.15cm} \\ \vspace{0.2cm} 
{\LARGE Michael A. Beasley$^1$$^,$$^4$} \\
$^1$ Centre for Astrophysics \& Supercomputing, Swinburne University, 
	Mail\#31, P.O.Box 218, Hawthorn, Vic 3122, Australia; \\
        {\rm \texttt{hclee@astro.swin.edu.au}}\\
$^2$ Center for Space Astrophysics, Yonsei University, 
	134 Shinchon, Seodaemoon, Seoul, 120-749, South Korea; \\
$^3$ Tuorla Observatory, V\"ais\"al\"antie 20, FIN-21500, 
	Piikki\"o, Finland; \\
$^4$ Lick Observatory, University of California, Santa Cruz, CA 95064, USA \\
}

\begin{document}
\date{\fbox{{\sc MNRAS, in preparation}: \today}}

\label{firstpage}

\maketitle

\begin{abstract}

The rotation curves of low surface brightness (LSB) galaxies suggest that they
possess significantly higher mass-to-light $(M/L)$ ratios than their high
surface brightness counterparts, indicating that LSB galaxies may be dark
matter dominated. This interpretation is hampered by the difficulty of
disentangling the disc and dark halo contributions from the disc dynamics of
LSB galaxies. Recently, Fuchs (2002) has attempted such a disentanglement using
spiral arm density wave and swing amplification theory, allowing an independent
measurement of the disc mass; this work suggests that LSB {\it discs} are
significantly more massive than previously believed. This would considerably
reduce the amount of matter required in the dark halos in fitting the rotation
curves. Interestingly, the high mass-to-light ratios derived for the discs
appear inconsistent with standard stellar population synthesis models.

In this paper, we investigate whether the high $M/L$ ratios for the Fuchs LSB
discs might be understood by adopting a very ``bottom heavy'' initial mass
function (IMF). We find that an IMF with a power law exponent of around 
$\alpha=3.85$
(compared to the standard Salpeter IMF, $\alpha=2.35$) 
is sufficient to explain the
unusually high $M/L$ ratios of the Fuchs sample.  Within the context of the
models, the blue colours ($(B-R)_0 < 1.0$) of the sample galaxies result from
being metal-poor ([Fe/H] = $-1.5 \sim -1.0$) and having undergone recent
($\sim$ 1 - 3 Gyr ago) star formation.

\end{abstract}


\section{Introduction}

Low surface brightness (LSB) galaxies have a wide range of properties, being
termed ``LSB'' only by virtue of their extrapolated central surface brightness
being fainter than $\mu_B$(0) = 21.65 $\pm$ 0.30 mag arcsec$^{-2}$
(Freeman 1970; see Bothun, Impey and McGaugh 1997; Impey and Bothun 1997 for
more detailed review of LSB galaxies).

Intriguingly, LSB galaxies appear, in general, to possess higher mass-to-light
$(M/L)$ ratios than their high surface brightness (HSB) counterparts. However,
the nature of this greater mass contribution is poorly constrained due to
uncertainties involved in the decomposition of baryonic disc matter (stars +
gas) from the non-baryonic dark matter halo, from disc dynamics (e.g., de Blok,
McGaugh and Rubin 2001).  Therefore, the total baryonic mass in the disc of LSB
galaxies, and its contribution to the disc dynamics, are not clear.

Recently, de Blok et al. (2001) have fitted a series of mass models to the
high-resolution rotation curves of 30 LSB galaxies, assuming both core
dominated (pseudo-isothermal) and cold dark matter cusp dominated 
(specifically,Navarro, Frenk and White 1997 (hereafter NFW) profiles) 
halos for a range of
assumptions about the stellar $M/L$ ratios.  They found that in general, the
pseudo-isothermal models fit the data better than the NFW profiles, and that the
maximum disc models are preferable for the fitting the inner rotation curves,
albeit requiring high stellar $M/L$ ratios.

Recently, a new constraint on the mass decomposition of LSB galaxy discs has
been made by Fuchs (2002) who employs the density wave and swing amplification
theory of galactic spiral arms.  The technique leads to a disc mass estimate
independent of traditional disc/halo decomposition of the rotation curve. He
reports that for his sample of LSB disc galaxies, selected for clear spiral
structure, the discs possess unexpectedly high $M/L$ ratios in the R-band
($M/L_R \sim 4 - 16$)\footnote{We have recalculated $M/L_R$ after removing the
gas mass component and listed them in Table 1}, in addition to relatively blue
colours ($(B-R)_0 < 1.0$).  Fuchs also found that the LSBs with higher $M/L$
ratios have lower fractions of gas to total baryonic mass, perhaps indicating
these galaxies have had a higher conversion rate of gas into stars.  Further,
the LSB discs for which a disc mass can be independently estimated, are
significantly heavier than expected for standard stellar populations synthesis
models (Bell and de Jong 2001; Mouhcine and Lancon 2003).

We investigate in this paper whether the high $M/L$ ratios of the Fuchs sample
can be explained by an increased contribution of low-mass stars in LSB galaxy
stellar populations, i.e. through modifying the initial mass function (IMF).

\section{Data} \label{sec:data}

Table 1 lists the seven LSBs from the Fuchs (2002) sample. 
The de-reddened $B-R$ colours
are from Table 2 of de Blok, van der Hulst and Bothun (1995)\footnote{$M_T$ in
their column 8, which is corrected for Galactic extinction, is used to obtain
colours.}. We note that McGaugh and de Blok (1997, hereafter MdB97) 
also used $M_T$ for their compilation of $B-V$ and $B-I$ colours,  
and we also list these colours in Table 1.

\begin{table*}
\caption[LSB galaxies from the Fuchs (2002) sample.]
{The Fuchs (2002) sample of LSB galaxies.}

\label{tab-1}
\begin{tabular}{lccccc}
\\\hline
Galaxy ID & $(B-R)_0$ & $(B-V)_0$ & $(B-I)_0$ & $M/L_R$ & $\mu_0$ \\ \hline\hline
F568$-$1  & 0.79  & 0.62  & 1.32  & 13.0 & 23.76 \\
F568$-$3  & 0.84  & 0.55  & 1.29  &  6.2 & 23.07 \\
F568$-$6  & -     & 0.63  & 2.26  & 10.3 & 23.60 \\
F568$-$V1 & 0.78  & 0.51  & 1.21  & 14.4 & 23.29 \\
UGC 128   & 0.69  & 0.51  & 1.09  &  3.1 & 24.22 \\
UGC 1230  & 0.76  & 0.52  & 1.20  &  3.0 & 23.36 \\
UGC 6614  & -     & 0.72  & 2.32  &  5.8 & 24.85 \\
\hline
\end{tabular}
\end{table*}

\begin{figure}  
\begin{center}
  \leavevmode
  \epsfxsize 8cm
  \epsfysize 8cm
  \epsffile{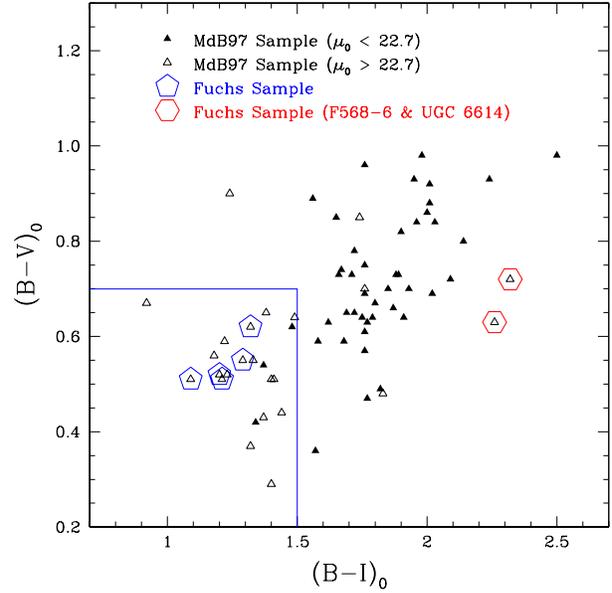}
\end{center}
\caption{Colour-colour diagram of the spiral galaxies from the sample of
McGaugh and de Blok (1997, MdB97). Five of the Fuchs (2002) sample of 
LSB galaxies are located at the blue end of the colour sequence (pentagons). 
The exceptions are F568-6 and UGC 6614 (hexagons; see text).}
\end{figure}

In Figure 1, we have divided the MdB97 sample of (both HSB 
and LSB) spiral galaxies into two
groups, one group with $\mu_{0}$ $<$ 22.7 and the other with $\mu_{0}$ $>$
22.7. This is a rather loose application of Freeman's law on this sample of
spiral galaxies. The location of the Fuchs sample in Figure 1 shows 
that his sample is, in general, fairly blue,
when compared to the MdB97 sample of spiral galaxies. 
The galaxies F568-6 (Malin2) and UGC 6614 are clear outliers in
$B-I$ colour.  These two are amongst the largest LSBs known (Quillen and
Pickering 1997).

\section{Stellar population models} \label{sec:model}

The present stellar population synthesis models are based upon the $Y^{2}$
Isochrones\footnote{http://csaweb.yonsei.ac.kr/$\sim$kim/yyiso.html} which are
computed until the onset of helium core burning (Kim et al. 2002).  
The models we compute are all single star bursts, characterised by a 
single age and metallicity. 
In reality, galaxies are generally composed of mixtures
of stellar populations, with a range of ages and 
metallicities. We have deliberately used a simplified (i.e., single burst)
star formation prescription for the LSB galaxies, in order to 
explore the importance of varying one parameter, namely
the IMF. In this regard, our single-burst models 
approximately represent the luminosity weighted 
mean age and metallicity for the stellar systems in question. 


Following our previous models (Lee et al. 2000, 2002),
the isochrones assume [$\alpha$/Fe] $= +0.3$, and are coupled to
the post-RGB stellar evolutionary tracks of Yi, Demarque and Kim (1997).
Although the latter tracks are solar-scaled, 
the effects of alpha-enhancements are
minimal at low-metallicity ($Z < Z_{\sun}$) (Kim et al. 2002) which is the
metallicity range of interest for our LSBs (see Section 4).  Furthermore, since
we employ steeper IMFs than the standard Salpeter one, the importance of
post-RGB stars is reduced because of the reduced number of high mass stars.

The investigated age range is from 1 Gyr to 14 Gyr, and the metallicities cover
$-2.51 \leq$ [Fe/H] $\leq +0.39$. The colours of the horizontal-branch have
been calibrated to Milky Way globular clusters (Lee et al. 2000, 2002).  The
stellar library of Lejeune, Cuisinier and Buser (1997, 1998) was used for the
conversion from theoretical to observable quantities because of their wide
coverage in stellar parameters such as temperature, surface gravity, and
metallicity.

 The constancy and shape of IMF is still controversial, 
although a number of studies are converging on 
the idea of a so-called `universal' IMF 
(e.g., Kroupa 2002; Wyse et al. 2002; Bell et al. 2003; 
Chabrier 2003; Weidner and Kroupa 2004). Indeed, the possible dependence
of IMF on galaxy types has been suggested by the previous studies
(e.g. Portinari et al. 2004). In this paper, we explore the 
variation in IMF required to explain the high M/L ratios
of LSB galaxies.
To this end, we use the simple power law IMF, where
the number of stars ($d N$) in mass interval ($dm$) is described by 
\begin{equation}
d N \propto m^{-\alpha} dm,
\end{equation}
The range in $\alpha$ required to
reproduce the unusually high $M/L$ ratios of LSB galaxies is 
shown in the Figures 2 to 4. 
For simplicity, the lower-mass (${\rm M_l}$) and the upper-mass
(${\rm M_u}$) ends of the IMF were fixed at 0.1$M_{\sun}$ and 60$M_{\sun}$,
respectively. 

In this paper, we define the IMF with 
$(\alpha,{\rm M_u},{\rm M_l})=(2.35,0.1,60)$
as the standard Salpeter model, although
the original Salpeter (1955) IMF is only defined 
for the mass function of stars within the mass range 
0.4--10 ${\rm M_{\sun}}$.
Due to this simplification, there is only one free parameter, $\alpha$,
to describe the shape of the IMF. This enables us to explore
other parameters, such as metallicity and age, more easily.
Of course, a more realistic IMF may be of a more complicated form, 
but this power law formula provides 
an approximation for the such  IMFs.

The low-mass end of the $Y^{2}$ isochrones is given at 0.4$M_{\sun}$ because
of the uncertain nature of the high density equation of state for
 low mass stars.
For our purposes, we linearly extrapolated the
$Y^{2}$ isochrones down to 0.1$M_{\sun}$\footnote{
Stellar models which reach masses below 
0.4 ${\rm M_{\sun}}$ have been presented in the literature
(e.g. Baraffe et al.\ 1998; Chabrier et al.\ 2000).
We use a simple extrapolation rather than these
more sophisticated models, but our conclusions should not be much affected
by this simplification. This is because the luminosity contribution
of these low mass stars is negligible, especially in
the optical band where we focus in this study. 
From our calculations, for instance, at [Fe/H] = -1.5
and at 3 Gyr, the mass fraction from the stars below 0.4 ${\rm M_{\sun}}$
is 87.2 \% while their contribution to the total 
luminosity is only 7.6 \%, in the case of $\alpha$ = 3.35.
}.
For IMFs steeper than the standard Salpeter variety, 
the baryonic $M/L$ ratios significantly depend upon the
choice of the low-mass end. For instance, for $M/L\sim$10, 
$\alpha$ + 4.5 is needed when the low-mass cut-off is taken at
0.4$M_{\sun}$, while $\alpha$ + 1.5 is necessary if the low-mass cut-off is
taken at 0.1$M_{\sun}$.

\section{RESULTS AND DISCUSSION} \label{sec:rad}

Figures 2 to 4 show our derived $M/L$ ratios and colours for single burst
populations for a range of galaxy ages and metallicities.  There is a region of
parameter space where a slightly steeper than normal (bottom-heavy) IMF does
achieve agreement with the unusually high $M/L_R$ ratios of the Fuchs sample.

The blue colours of the LSBs also provide constraints
upon the allowable combination of age and metallicity of the sample. The
characteristic colours of $(B-R)_0 \leq 1.0$, for instance, require recent
star formation ($\sim$ 1-3 Gyr ago), when adopting the metallicity range 
--1.5$<$[Fe/H]$<$--1.0, as implied by de Blok \& van der Hulst (1998)'s 
oxygen abundance measurements.  
This interpretation for such blue colours is generally
consistent with the recent study by Burkholder, Impey and Sprayberry (2001).
Interestingly, Padoan, Jimenez and Antonuccio-Delogu (1997) have
concluded that LSB galaxies are typically older than 7 Gyr. 
Their conclusions are, however, only valid 
where the metallicity of LSB galaxies are constrained to Z = 0.0002 ([Fe/H]
$\sim$ $-$2.0) as they stressed in their paper. This is a result of
the well-known age-metallicity degeneracy  which plagues 
integrated broadband colours in the optical (e.g., Worthey 1994).

\begin{figure}  
\begin{center}
  \leavevmode
  \epsfxsize 8cm
  \epsfysize 9cm
  \epsffile{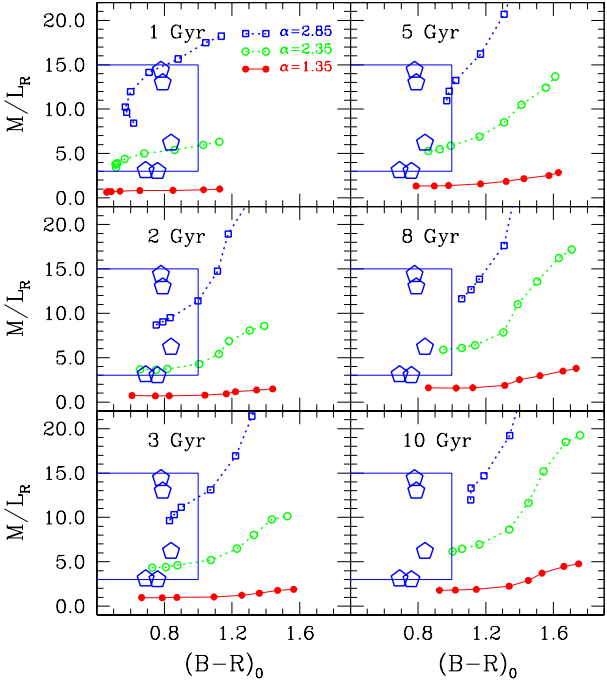}
\end{center}
\caption{Our simple stellar population predictions for the $(B-R)_0$
vs. $M/L_R$ as a function of age for three IMF slopes. The solid lines with
filled circles correspond to the standard Salpeter IMF with an
exponent $\alpha=2.35$. The dotted lines with open circles and open squares are
$\alpha=3.35$ and $\alpha=3.85$ cases, respectively. At a given age and IMF
exponent, the metallicity [Fe/H] is $-2.51$, $-1.90$, $-1.51$, $-0.90$,
$-0.49$, $-0.17$, $+0.17$, and $+0.39$, from left to right.  From the left
panel, it is suggested that both unusually high $M/L_R$ ratios and blue colours
($(B-R)_0$ $<$ 1.0) of the Fuchs (2002) sample of LSB galaxies (large pentagons)
could be satisfactorily explained by employing steeper than Salpeter IMFs with
the combination of low metallicity ([Fe/H] = $-$1.5 $\sim$ $-$1.0) and a rather
recent ($\sim$ 1 - 3 Gyr) star formation.}
\end{figure}

\begin{figure}  
\begin{center}
  \leavevmode
  \epsfxsize 8cm
  \epsfysize 9cm
  \epsffile{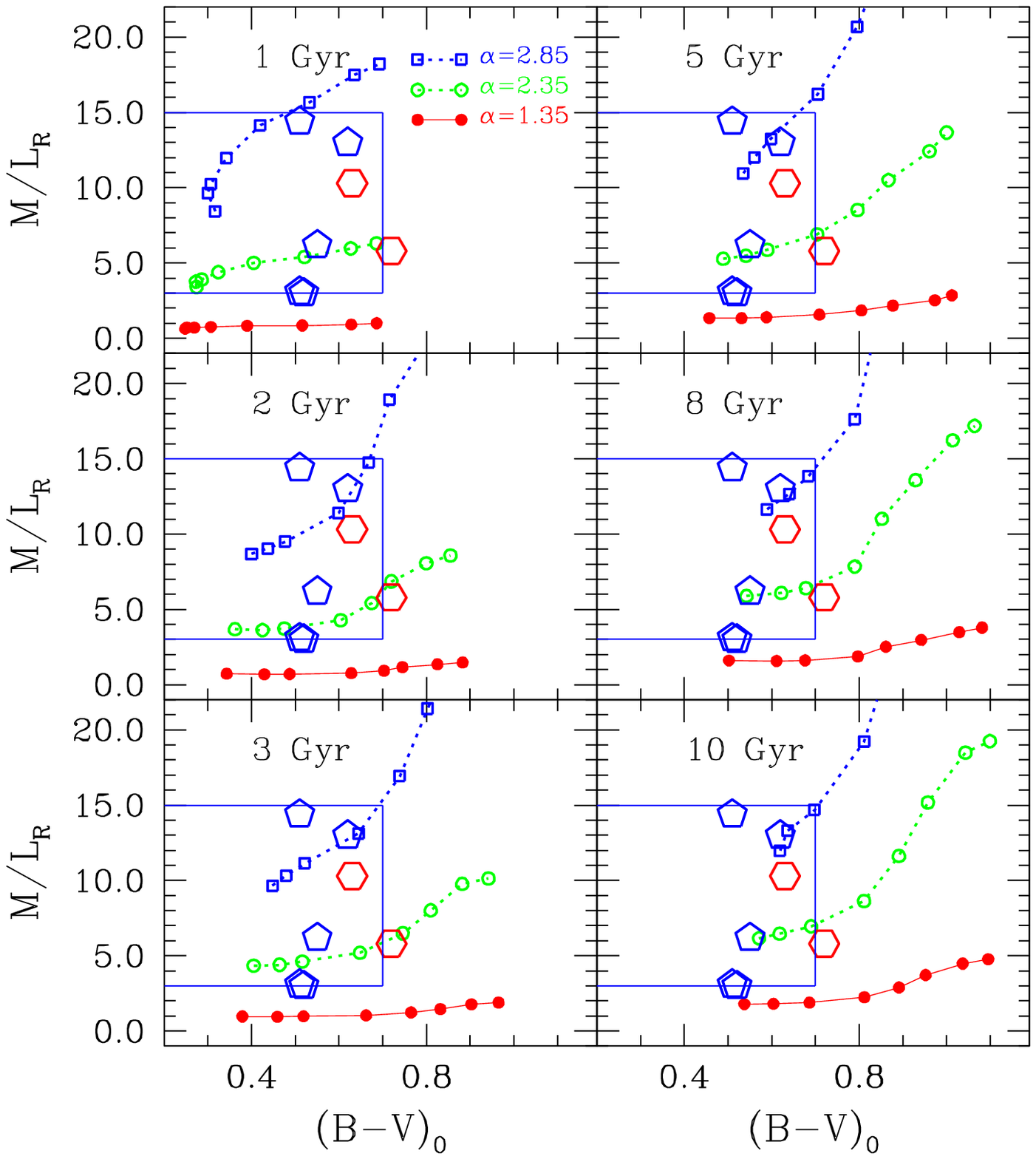}
\end{center}
\caption{Similar to Figure 2, but $(B-V)_0$ vs. $M/L_R$ diagrams.  The two large
hexagons are F568-6 ($M/L_R=10.3$) and UGC 6614 ($M/L_R=5.8$).}
\end{figure}

\begin{figure}  
\begin{center}
  \leavevmode
  \epsfxsize 8cm
  \epsfysize 9cm
  \epsffile{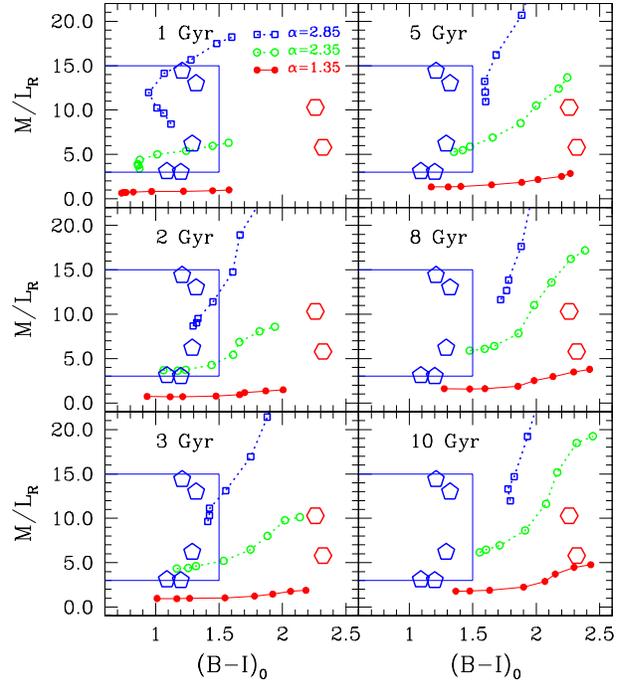}
\end{center}
\caption{Similar to Figure 2, but $(B-I)_0$ vs. $M/L_R$ diagrams.  The two large
hexagons are F568-6 ($M/L_R$=10.3) and UGC 6614 ($M/L_R$=5.8).
}
\end{figure}

Figure 5 shows that our simple stellar population models broadly follow the
colours of the MdB97 sample (both HSB and LSB spiral galaxies) 
in the $(B-I)_0$ vs. $(B-V)_0$ plane. 
In order to explain their presently rather blue colours, Figures 2 - 5 suggest 
that the LSBs do need the combination of low 
metallicity ([Fe/H] = $-$1.5 $\sim$ $-$1.0) and 
rather recent star formation ($\sim$ 1 $-$ 3 Gyr ago). The two giant LSBs
(F568-6 and UGC 6614) almost certainly require an additional older, metal rich
population to explain their red colours.

A notable feature in Figures 2 to 5 is that the integrated colours at
1 Gyr are significantly bluer than those at 2 Gyr.  This is because
of the convective core overshooting phenomenon, which causes a brighter
main-sequence turnoff and a longer stellar lifetime during the hydrogen burning
phase (Yi 2003).

\begin{figure}  
\begin{center}
  \leavevmode
  \epsfxsize 8cm
  \epsfysize 10cm
  \epsffile{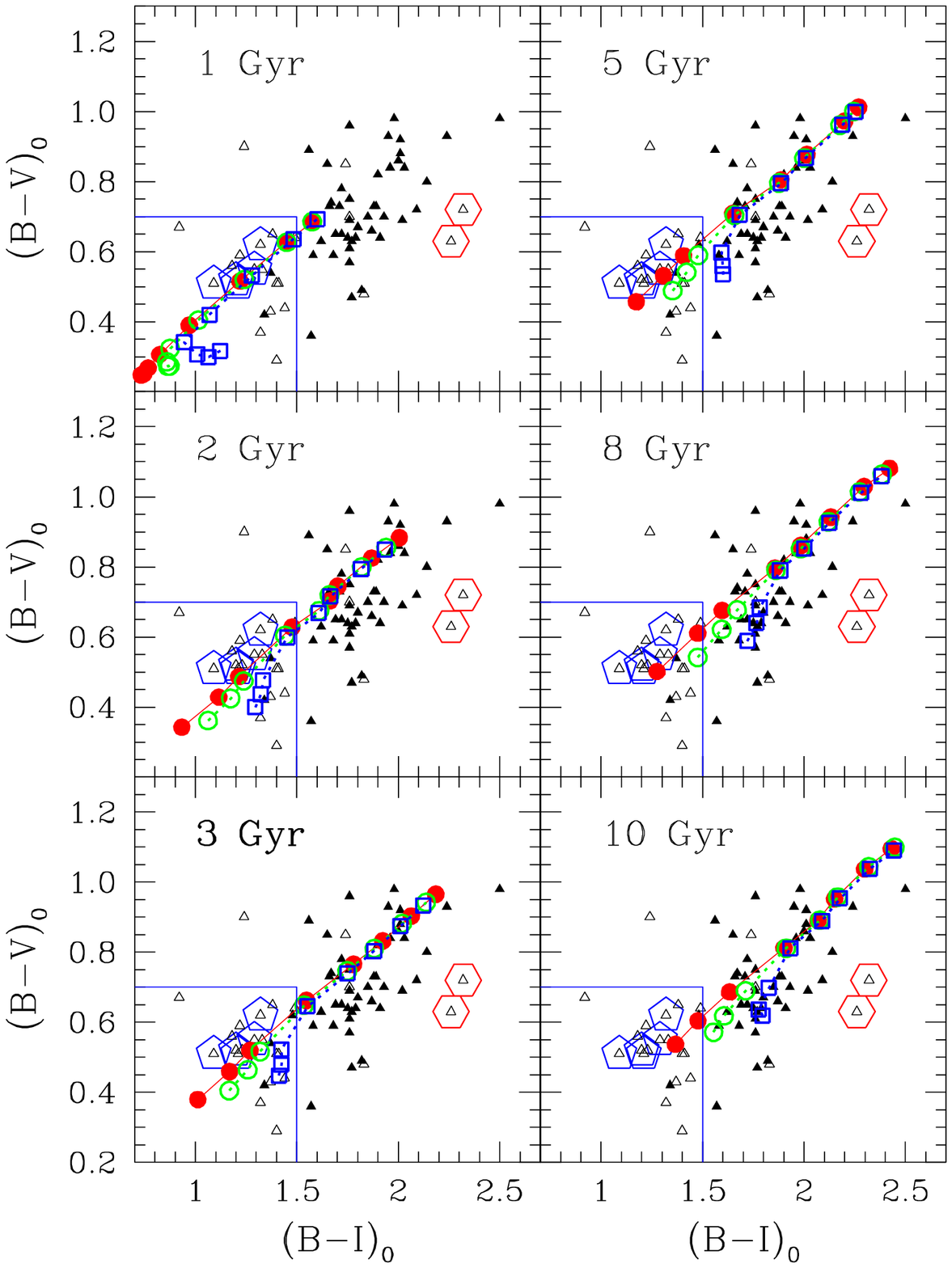}
\end{center}
\caption{ Our simple stellar population predictions are compared with the
sample of MdB97 (see Figure 1 for symbol description).  The solid line with
filled circles is for standard Salpeter IMF with $\alpha=2.35$. The dotted lines
with open circles and open squares are $\alpha=3.35$ and $\alpha=3.85$ cases,
respectively.  At the given IMF exponent, [Fe/H] is $-$2.51, $-$1.90, $-$1.51,
$-$0.90, $-$0.49, $-$0.17, +0.17, and +0.39 from lower left to upper right.  }
\end{figure}

\subsection{IMF variations}

The adoption of a steeper-than-Salpeter IMF is easily able to explain the
$M/L$ ratios of the sample LSB disc galaxies. Is such a steep IMF plausible?

From the observed CO deficiency in LSB galaxies, Schombert et al. (1990) have
suggested that star formation may be inefficient in metal-poor discs, and may
take place preferentially outside the giant molecular cloud environment with an
IMF operating in a suppressed high mass mode.  It is interesting to note, in
this respect, that the Romanishin, Strom and Strom (1983)'s spectroscopy of a
few nearby LSB galaxies suggests that their H II regions are significantly
deficient in massive stars.  Schombert et al. (1990) further proposed that the star
formation that takes place in warmer, low-density H I environments could
produce smaller clouds (because of initially lower cloud masses) and have an
IMF which is substantially different to that in an environment of cold, dense
gas.  Interestingly, Waller et al. (2002) find from multi-band HST
photometry of H II regions in M33 that the IMF may depend on metallicity in the
sense that the most metal-rich H II regions produce more massive than low-mass
stars. This is a rather surprising result when one considers the 
generally favoured top-heavy IMF theory at early times when metallicity is 
low (e.g., Larson 1998).

If the stellar IMF of the LSB galaxies is indeed steeper than the standard
Salpeter IMF, then this might imply that (1) the LSBs must be intrinsically
less luminous in the optical as well as in the near-IR regime compared
to HSB galaxies. This simply due to the significantly smaller number fraction 
of luminous stars in LSB galaxies, despite their similar masses to
HSBs (2) this smaller number fraction of evolved stars leads
naturally to less efficient chemical enrichment, and thus lower metallicity (3)
in such a low metallicity environment, less efficient dust production may
further accentuate the bluer colours of LSBs, and (4) significant amounts of
baryonic matter, in the form of low-mass stars, may reside in the discs of LSB
galaxies. Depending upon the shape of the LSB luminosity function, LSBs could
be major repositories of baryons in the Universe.
 
Another interesting issue is the gas mass fraction.  LSB disc galaxies are
known to be H I rich relative to their optical luminosity, when
compared to HSB galaxies. The baryonic gas
mass fraction is written

\begin{equation}
f_g = {M_g \over {M_g + M_*}}
\end{equation}

where $M_g$ is the total mass in the form of gas, and $M_*$ is the mass in
stars.  The estimation of $M_*$ is, however, now arguably quite uncertain.  The
$f_g$ therefore depends upon what amounts of stellar mass one adopts, and
admittedly the adoption of the standard Salpeter IMF for the $M_*$ has been
prevalent in the literature.  Like our case of a steeper than the standard
Salpeter IMF, if the stellar mass is indeed significantly enhanced by low-mass
stars and more of the gas is locked up into these stars, then $f_g$ should
become correspondingly smaller.  A hint toward this interpretation is seen in
figure 5 of Fuchs (2002).  Accordingly, if the steeper than Salpeter IMF is
truly the case for the LSB galaxies, then star formation scenarios
at lower efficiencies may need to be reconsidered as a result of this 
study (e.g., van den Hoek et al. 2000).

\section{SUMMARY} \label{sec:sum}

We have presented theoretical efforts to reproduce the
relatively high mass-to-light ratios, and blue colours, of 
a sample of LSB galaxies by employing a significantly 
steeper IMF than the standard Salpeter variety. 
In particular, we address the
unusually high M/L ratios that are based upon Fuchs' independent disc mass
estimation of several LSB disc galaxies with clear spiral structure, using his
density wave and swing amplification theory.  We find that a ``bottom-heavy''
IMF, such as $\alpha$ = 3.85, is sufficient to reproduce the
high M/L ratios. Moreover, we report that a combination of a relatively low
metallicity ([Fe/H] = $-$1.5 $\sim$ $-$1.0) and a rather recent ($\sim$ 1 - 3
Gyr ago) star formation could also successfully explain the observed blue
colours of the LSB galaxies ($(B-R)_0$ $<$ 1.0).  
Although suggestive, it remains to be seen whether these simple
IMF variations are justifiable, or whether more complicated processes are
needed. 

\section*{Acknowledgements}

It is a pleasure to thank Burkhard Fuchs who pointed out this 
puzzling but intriguing realm of LSB galaxies.
We also thank Erwin de Blok, Myung-Hyun Rhee and Aeree Chung for useful
discussions.  
We thank the anonymous referee for her/his expedient reading of this paper
and helpful comments.
H.-c. Lee acknowledges the financial support of the Australian
Research Council Linkage International Fellowship and the Creative 
Research Initiatives Program of the Korean Ministry of Science and 
Technology.

\end{document}